\documentstyle[epsfig]{europhys}

\def\etal{{\hbox{{\it\ et al.\/}\rm :\ }}}

\def\stars{\bigskip\centerline{***}\medskip}

\newif\ifboo \boofalse

\begin{document}

\euro{}{}{}{}

\Date{29 July 1998}

\shorttitle{M. M. S\'ANCHEZ \etal THE HALF-FILLED HUBBARD CHAIN
IN COM}

\title{The half-filled Hubbard chain in the Composite Operator
Method: A comparison with Bethe Ansatz}

\author{M. M. S\'anchez\thanks{E-mail: sanmar@vaxsa.csied.unisa.it},
A. Avella and F. Mancini}

\institute{Universit\`a degli Studi di Salerno -- Unit\`a INFM
di Salerno\\ Dipartimento di Scienze Fisiche ``E. R. Caianiello''\\
84081 Baronissi, Salerno, Italy}

\pacs{
\Pacs{71}{10.$-$w}{Theories and models of many electron systems}
\Pacs{71}{10.Fd}{Lattice fermion models}
\Pacs{71}{15.Th}{Other methods}}

\rec{29 July 1998}{}

\maketitle

\begin{abstract}
The one-dimensional Hubbard model at half-filling is studied in the
framework of the Composite Operator Method using a static approximation.
A solution characterized by strong antiferromagnetic correlations and a
gap for any nonzero on-site interaction $U$ is found. The corresponding
ground-state energy, double occupancy and specific heat are in excellent
agreement with those obtained within the Bethe Ansatz. These results
show that the Composite Operator Method is an appropriate framework for
the half-filled Hubbard chain and can be applied to evaluate properties,
like the correlation functions, which cannot be obtained by means of the
Bethe Ansatz, except for some limiting cases.
\end{abstract}

\section{Introduction}
New materials whose physics is dominated by electron correlations in
narrow energy bands are actually a challenge for solid state physicists.
The treatment of such systems is not trivial due to the competition
between itinerant and localized behaviour of the electrons in these
bands. The Hubbard Hamiltonian \cite{hubbard} is regarded as the
simplest model which can give us the basic understanding of the effects
of strong electronic correlations. In particular, its one-dimensional
(1D) version is interesting for several reasons. On the one hand it is
exactly integrable, by means of the Bethe Ansatz \cite{bethe}. On the
other, it could be applied to study real quasi-1D systems like the $KCP$
and $TTF-TCNQ$ salts \cite{epstein}, the doped spin Peierls chains
\cite{book1D} and the Cu-O chains of the high-$T_c$ superconductors,
whose underlying physics is directly related to the low-dimensionality
of the system.

By using the Bethe Ansatz and following a method developed by Yang
\cite{yang}, Lieb and Wu evaluated exactly some ground-state properties
of the 1D Hubbard model at half filling \cite{liebwu}. Two years later,
the spin and charge excitation spectra were obtained, also within the
Bethe Ansatz, by Ovchinnikov \cite{ovchinnikov}. For arbitrary electron
density, the Bethe Ansatz coupled integral equations for the charge and
spin distribution functions cannot be analytically solved except in some
limiting cases \cite{carmelo88} and numerical calculation is needed. In
this way, Shiba \cite{shiba} evaluated the ground state energy, local
magnetic moment and magnetic susceptibility as a function of the
electron density. The finite-temperature formalism of the Bethe Ansatz
developed by Takahashi \cite{takahashi} provides the evaluation of the
thermodynamic properties \cite{usuki}. Further studies of the 1D Hubbard
model within the framework of the Bethe Ansatz have clarified for
instance the thermodynamics in the presence of a magnetic field
\cite{carmelo91}.

Therefore, we see that the Bethe Ansatz allows to exactly evaluate many
quantities, thus providing a wide picture of the physics of the Hubbard
chain. Nevertheless, this Ansatz cannot be regarded as a complete
framework since many important properties, like the charge and spin
correlation functions and the spectral properties, cannot be extracted
from the exact Bethe Ansatz wave function except for some limiting cases
(e.g. half filling, $U\rightarrow \infty$, $\omega=0$) \cite{ogata}.
Otherwise we must address to other numerical \cite{num} or analytical
\cite{anal} approaches.

The Bethe Ansatz is a very useful test for any approximation scheme,
whose reliability can be checked by computing quantities that are given
exactly by such Ansatz. In this sense, many of the available analytical
methods do not give satisfactory results. On the other hand, all
numerical techniques present some inherent problems, namely the small
size of the clusters and the impossibility of reaching very low
temperatures. Motivated by this, we have studied the 1D Hubbard model at
half filling by means of an analytic approach that has proved to be
adequate for studying other strongly correlated models \cite{COM}. The
results obtained for the thermodynamic properties are in good agreement
with the numerical data \cite{2Dthermo}, also some anomalous
thermodynamic and magnetic behaviours observed in high-$T_{c}$ cuprate
superconductors have been successfully explained \cite{2Dthermo,incom}.
In this calculation scheme, called the Composite Operator Method (COM),
the long-lived excitations of the system are described by an appropriate
combination of the standard fermionic field operators. The properties of
the new fermionic fields are self-consistently determined by the
dynamics. To fix the {\em internal} parameters that appear, some
symmetry requirements, like the Pauli principle and the particle-hole
symmetry, are imposed. This procedure permits to recover symmetries that
are badly violated by other approaches \cite{rowe}, and thus is expected
to provide a better description of strongly correlated systems
\cite{2pole}.

In this letter we analyze the half-filled infinite Hubbard chain within
COM in the static approximation, where finite life-time effects are
neglected. Although we are considering a paramagnetic (PM) ground state,
we find a solution of the model which shows strong antiferromagnetic
(AF) features. We present a detailed analysis of this AF-like solution.
We calculate the energy and double occupancy of the ground-state, and
the specific heat as a function of temperature, and compare them to the
exact results obtained by the Bethe Ansatz. Excellent agreement is
found.

\section{Method}
We consider the well-known Hubbard Hamiltonian:
\begin{equation}
H=\sum_{ij}t_{ij}c^{\dagger }\left( i\right) c\left( j\right)
+U\sum_{i}n_{\uparrow }\left( i\right) n_{\downarrow }\left( i\right)
-\mu\sum_{i}n(i)
\end{equation}
where $c^{\dagger }\left( i\right) =\left( c_{\uparrow }^{\dagger
}\left( i\right) ,c_{\downarrow }^{\dagger }\left( i\right) \right) $ is
the electron operator on the site $i$ in the spinor notation, $n_{\sigma
}\left( i\right)$ is the charge--density operator for the spin $\sigma$,
$\mu$ is the chemical potential introduced to control the particle
density $n=\langle c^\dagger(i)c(i)\rangle $, and $U$ is the on-site
Coulomb interaction. Considering only nearest neighbours and taking the
lattice constant as one, the hopping matrix for the chain is
\begin{equation}
t_{ij}=-2t\frac{1}{N}\sum_{k}e^{ik\left( i-j\right)}cos(k)
\end{equation}
In the case of the Hubbard model, a natural choice for the composite
operators is the Hubbard doublet
$\Psi^\dagger=(\xi^\dagger(i),\eta^\dagger(i))$, where
\begin{eqnarray}
\xi^\dagger(i)&=&c^\dagger(i)\,[1-n(i)]\nonumber\\
\eta^\dagger(i)&=&c^\dagger(i)\,n(i)
\end{eqnarray}
This operators describe the hopping of an electron to an unoccupied and
to an occupied site $i$, respectively. Considering a two-pole
approximation \cite{Npole} and a PM ground state, the Fourier transform
of the single-particle retarded thermal Green's function may be written
in {\em COM} as:
\begin{equation}
S\left( k,\omega \right) =\sum_{i=1}^{2}\frac{\sigma ^{(i)}
\left( k\right) }{\omega -E_{i}\left( k\right) }
\label{GF}
\end{equation}
The energy bands are given by $E_i(k)=R(k)+(-)^{i+1}\,Q(k)$ where
\begin{eqnarray}
R(k)&=&\frac{U}{2}-\mu-2t\,cos(k)-\frac{m(k)}{2\,I_1I_2}\nonumber\\
Q(k)&=&\frac{1}{2}\sqrt{\left(U-\frac{m(k)}{I_1I_2}\right )^2+2nU\,
\frac{m(k)}{I_1I_2}} \label{bands}\\
m(k)&=&2t[\Delta+\cos(k)\,(p-I_2)]\nonumber
\end{eqnarray}
and $I_1$, $I_2$ are the diagonal matrix elements of the normalization
matrix $I
= F.T.\left\langle\left\{\Psi(i),\Psi^\dagger(j)\right\}\right\rangle$.
The spectral moments $\sigma^{(i)}(k)$ that appear in the Green's
function (\ref{GF}) are given in ref. \cite{COM,Npole}.

As shown above, the single-particle thermal Green's function (\ref{GF})
depends on the {\em external} parameters $t$, $U$, $n$ and $T$
(temperature), and three {\em internal} parameters $\mu$, $\Delta$ and
$p$. $\Delta$ and $p$ are the following intersite charge correlation
function and intersite charge, spin and pair correlation function,
respectively \cite{COM,2pole}.
\begin{eqnarray}
\Delta&=&\langle\xi^\alpha(i)\xi^\dagger(i)\rangle-
\langle\eta^\alpha(i)\eta^\dagger(i)\rangle\nonumber\\
p&=&\frac{1}{4}\langle n_\mu^\alpha(i)n_\mu(i)\rangle -\langle
(c_\uparrow(i)
c_\downarrow(i))^\alpha\,c_\downarrow^\dagger(i)c_\uparrow^\dagger(i)\rangle
\end{eqnarray}
The superscript $\alpha$ indicates the field on the first neighbour
sites and $n_\mu(i)=c^\dagger(i)\sigma_\mu c(i)$ is the charge-
($\mu=0$) and spin- ($\mu=1,2,3$) density operator, where $\sigma_\mu$
is the Pauli representation of the $SU(2)$ symmetry group. These
parameters roughly produce (see eq.~(\ref{bands})) a shift in the bands
($\Delta$) and a bandwidth renormalization ($p$). Very different results
are obtained according to how these internal parameters are fixed
\cite{2pole}. In {\em COM} they are determined by solving the following
system of coupled self-consistent equations,
\begin{equation}
\left\{
\begin{array}{l}
n=2\left( 1-S_{11}-S_{22}\right) \\
\Delta =S_{11}^{\alpha }-S_{22}^{\alpha } \\
S_{12}=0
\end{array}
\right.
\label{self}
\end{equation}
with
$S_{\gamma\delta}=\left\langle\Psi_\gamma(i)\Psi^\dagger_\delta(i)\right\rangle$
and
$S_{\gamma\delta}^\alpha=
\left\langle\Psi^\alpha_\gamma(i)\Psi^\dagger_\delta(i)\right\rangle$.

The first two equations are obtained from the existing relations with
the elements of the Green's function, and the third one comes from
requiring the satisfaction of the Pauli principle at the level of matrix
elements (see ref.~\cite{COM,2pole} for details).

Once the internal parameters are determined, the evaluation of the
physical quantities is straightforward. In this paper we study the
single-particle band structure, the energy and double occupancy of the
ground state, and the specific heat. The band structure is determined
from eq.~(\ref{bands})). The ground-state energy per site is calculated
as the thermal average of the Hamiltonian and is given by
\begin{equation}
E=2t\left(G_1-(1-n)UF_1+\frac{B_1}{I_1I_2}\right)+\frac{I_2}{2}U(1-G_0-UF_0)
\label{energy}
\end{equation}
where the functions $F_n$, $G_n$ and $B_n$ ($n=0,1,\cdots$) are defined
in ref.~\cite{2pole}.

We use the thermodynamic relations $E=F+TS$,
\begin{equation}
F(T,n)=\int_0^n\,\mu(T,n')\,dn'\hspace{1cm} S(T,n)=-\int_0^n\,
\left(\frac{\partial\mu}{\partial T}\right)_{n'}\,dn'
\end{equation}
with $F$ the Helmholtz free energy and $S$ the entropy, to determine the
specific heat, which reads as
\begin{equation}
C(T,n)=-T\int_0^n\, \left(\frac{\partial^2\mu}{\partial
T^2}\right)_{n'}\,dn'
\label{cv}
\end{equation}
As shown for the 2D case \cite{2Dthermo}, the temperature derivatives of
the chemical potential can be expressed in terms of the internal
parameters and calculated once the self-consistent equations
(\ref{self}) are solved. Finally, the double occupancy is calculated by
deriving the free energy with respect to $U$.

\section{Results}
\begin{figure}[tb]
\begin{center}
\epsfig{file=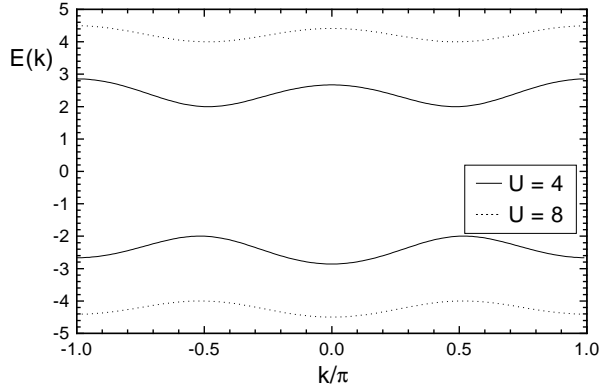,width=8cm,clip=}
\end{center}
\caption{Band structure of the half-filled Hubbard chain at $T = 0$ for
the indicated couplings.}
\label{fig1}
\end{figure}

Considering a PM ground state, we find a fully self-consistent solution
characterized by a negative value of the $p$ parameter. The results for
the energy spectrum are shown in fig.~\ref{fig1}. As we can see, this
solution presents a typical AF band pattern; namely, a first excitation
at $k=\pm\frac{\pi}{2}$, a very narrow bandwidth of order
$J=\frac{2t^2}{U}$ (the AF exchange interaction), and a quasi-halved
Brillouin zone. Such an AF-like band structure is directly related to
the negative sign of the $p$ parameter, which is responsible of the
general band shape [see eq.~(\ref{bands})]. In the figure, the energy is
measured with respect to the chemical potential. The solution exhibits a
gap in the excitation spectrum for any non-zero value of $U$, in
agreement with the Bethe Ansatz solution. The rate at which the gap
increases coincides with Bethe Ansatz down to $U\approx 4$.

\begin{figure}[tb]
\begin{center}
\epsfig{file=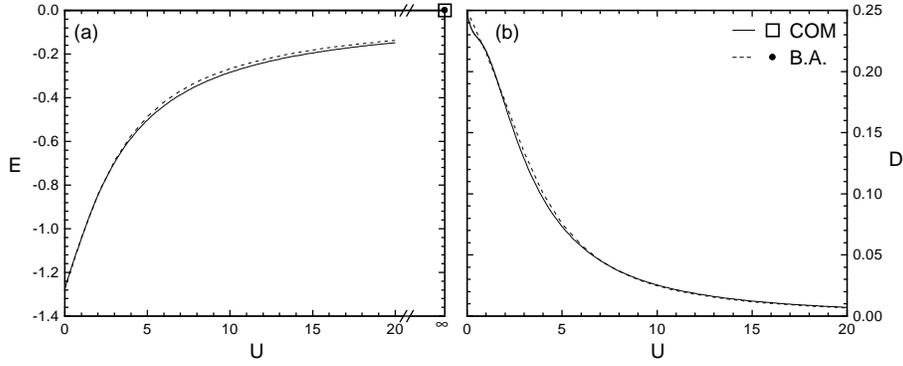,width=12cm,clip=}
\end{center}
\caption{$(a)$ Ground-state energy and $(b)$ double occupancy
of the half-filled Hubbard chain in {\it COM} (solid line) and Bethe's
Ansatz (dashed line).}
\label{fig2}
\end{figure}

\begin{figure}[tb]
\begin{center}
\epsfig{file=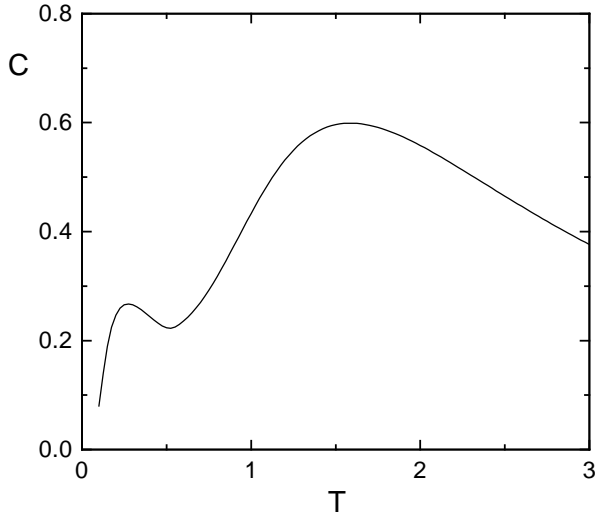,width=8cm,clip=}
\end{center}
\caption{Specific heat versus temperature of the half-filled Hubbard
chain in {\it COM} for $U = 8$.}
\label{fig3}
\end{figure}

We calculate the ground-state energy and double occupancy as indicated
in the previous section. These quantities, as a function of $U$, are
shown in fig.~\ref{fig2} together with Bethe's Ansatz. The excellent
agreement obtained is probably related to the opening of the gap
mentioned above. Such a good agreement with Bethe Ansatz is not reached
by other analytical approaches like the Gutzwiller approximations or the
ladder and self-consistent ladder approximation \cite{anal}. In
particular, these approaches fail to reproduce the correct asymptotic
behaviour of the ground-state energy. As shown in panel $(b)$, the
double occupancy goes to zero as $U\rightarrow\infty$. In a system of
itinerant electrons with an average of one electron per site, the
kinetic energy must also have this asymptotic behaviour, because any
electron hopping would lead to a double occupied site. Therefore,
electrons localize at infinite $U$, where the half-filled Hubbard chain
is equivalent to the spin-$\frac{1}{2}$ AF Heisenberg chain that
describes a system of localized spins.

We calculate the specific heat of the system by means of eq. (\ref{cv}).
The result as a function of temperature and for a ratio $\frac{U}{W}=2$
(coupling/bandwidth) is shown in fig.~\ref{fig3}. A double-peak
structure appears, with a first peak located at $T\approx 0.25$ and a
second one at $T\approx 1.6$ ($T$ is measured in units of $t$). This is
in good agreement with the Bethe Ansatz calculation of Ref.
\cite{usuki}. The low-$T$ feature is due to the spin excitations, since
$0.25$ is precisely the magnitude of the AF exchange parameter
$J=\frac{2t^2}{U}$ for $U=8$ and a peak at such location is also found
in the AF Heisenberg chain \cite{bonner}. The high-$T$ peak is
associated to charge excitations. We can reach this interpretation by
examining the chemical potential as a function of the electron density
for several temperatures. At very low $T$, $\mu$ has a discontinuity at
$n=1$ of magnitude $U$ that leads to the opening of a gap in the charge
excitation spectrum, as we commented above. For increasing $T$ the
discontinuity in $\mu$ is smoothed because the electrons can be excited
across the gap. Therefore, charge excitations appear in the system at
higher temperatures. Such a specific heat structure of the 1D Hubbard
model, with low- and high- $T$ regions dominated by spin and charge
excitations, respectively, is also obtained by numerical calculations on
finite chains \cite{shibapincus}. In the weak-intermediate-coupling
regime $\left(\frac{U}{W}\leq 1\right)$ charge and spin excitations
cannot be distinguished since they are of the same energy range, and
hence the specific heat shows only one peak \cite{shibapincus}.

\section{Conclusions}
Summarizing, we have studied the 1D Hubbard model at half filling by
means of the {\it COM} considering a PM state and a two-pole
approximation. Within this approach we find a solution of the model that
reproduces very well the exact solution given by the Bethe Ansatz.
Namely, we obtain an insulating ground state characterized by strong AF
correlations for any non-zero interaction $U$. The energy and double
occupancy of this AF-like ground state is in excellent agreement with
that of Bethe Ansatz. We have also calculated the temperature dependence
of the specific heat of the system in the strong coupling regime. The
locations of the spin- and charge-excitation peaks are consistent with
the ones of the Bethe Ansatz. The approximation considered seems thus to
be an adequate framework to study the physics of the half-filled Hubbard
chain. It is of particular interest to apply this approach to get
information about properties, like the correlation and spectral
functions, which cannot be extracted from the Bethe Ansatz, except for
some limits.

The good agreement obtained for the ground-state energy and double
occupancy within a static approximation is not surprising. As it is well
known, such an approach gives a good description of the high-energy
sector even though is unable to distinguish the low-energy features.
Anyway, these features, are swallowed when integrating over the whole
energy range. In this line of thinking, it is worth noticing the good
description that we obtain for both the low- and high- energy sector of
the specific heat. We suspect that this is due to our correct treatment
of the Pauli principle at the level of matrix elements. In a physics
dominated by strong electronic correlations, we believe that the
satisfaction of the Pauli principle not only at the operator level, but
also at the level of observation (relation among matrix elements) is
crucial.

\stars

The authors wish to thank F.~D.~Buzatu for many valuable discussions.
M.~M.~S\'{a}nchez acknowledges a grant from the {\it Instituto Nazionale
per la Fisica della Materia} (INFM).


\vskip-12pt
\begin{thebibliography}{20}

\bibitem{hubbard} J. Hubbard, {\it Proc. Roy. Soc. London A}, {\bf 276}
(1963) 238.

\bibitem{bethe} H. Bethe, {\it Z. Physik}, {\bf 71} (1931) 205.

\bibitem{epstein} D. J\'{e}rome
and A. J. Schulz, {\it Adv. Phys.}, {\bf 31} (1982) 299; C. S. Jacobsen,
I. Johannsen and K. Bechgaard, {\it Phys. Rev. Lett.}, {\bf 53} (1984)
194.

\bibitem{book1D} T. M. Rice, in {\it Physics in One Dimension}, edited
by J. Bernasconi and T. Schneider, Vol. {\bf 23} (Springer Series in
Solid-State Sciences) 1981, pp. 229-238.

\bibitem{yang} C. N. Yang, {\it Phys. Rev. Lett.}, {\bf 19} (1967) 1312.

\bibitem{liebwu} E. H. Lieb and F. Y. Wu, {\it Phys. Rev. Lett.}, {\bf
20} (1968) 1445.

\bibitem{ovchinnikov} A. A. Ovchinnikov, {\it Sov. Phys. JETP}, {\bf 30}
(1970) 1160.

\bibitem{carmelo88} J. Carmelo and D. Baeriswyl, {\it Phys. Rev. B}, {\bf 37}
(1988) 7541.

\bibitem{shiba} H. Shiba, {\it Phys. Rev. B}, {\bf 6} (1972) 930.

\bibitem{takahashi} M. Takahashi, {\it Prog. Theor. Phys.}, {\bf 47}
(1972) 69.

\bibitem{usuki} N. Kawakami, T. Usuki and A. Okiji, {\it Phys. Lett. A},
{\bf 137} (1989) 287; T. Usuki, N. Kawakami and A. Okiji, {\it Journ.
Phys. Society of Japan}, {\bf 59} (1990) 1357.

\bibitem{carmelo91} J. Carmelo, P. Horsch, P. A. Bares and A. A.
Ovchinnikov, {\it Phys. Rev. B}, {\bf 44} (1991) 9967.

\bibitem{ogata} M. Ogata and H. Shiba, {\it Phys. Rev. B},
{\bf 41} (1990) 2326; N. Kawakami and A. Okiji, {\it Phys. Rev. B}, {\bf
40} (1989) 7066; N. Kawakami and S.-K. Yang, {\it Phys. Rev. Lett.},
{\bf 65} (1990) 3063; H. J. Schulz, {\it Phys. Rev. Lett.}, {\bf 64}
(1990) 2831; S. Sorella and A. Parola, {\it J. Phys. Condens. Matter},
{\bf 4} (1992) 3589; B.S. Shastry and B. Sutherland, {\it Phys. Rev.
Lett.}, {\bf 65} (1990) 243.

\bibitem{num} S. Sorella {\it et al.}, {\it Europhys. Lett.}, {\bf 12}
(1990) 721; J. H. Xu and J. Yu, {\it Phys. Rev. B}, {\bf 45} (1992) 6931
; R. Preuss {\it et al.}, {\it Phys. Rev. Lett.}, {\bf 73} (1994) 732.

\bibitem{anal}  W. Metzner and D. Vollhardt, {\it Phys. Rev. Lett.},
{\bf 59} (1987) 121; F. D. Buzatu, {\it Mod. Phys. Lett. B}, {\bf 9}
(1995) 1149; J. E. Hirsch, {\it Phys. Rev. B}, {\bf 22} (1980) 5259.

\bibitem{COM} S. Ishihara {\it et al.}, {\it Phys. Rev. B}, {\bf 49} (1994)
1350; F. Mancini {\it et al.}, {\it Physica C}, {\bf 244} (1995) 49;
{\bf 250} (1995) 184; {\bf 252} (1995) 361; F. Mancini {\it et al.},
{\it Phys. Lett. A}, {\bf 210} (1996) 429; A. Avella {\it et al.}, {\it
Physica C}, {\bf 282--287} (1997) 1757; {\bf 282--287} (1997) 1759.

\bibitem{2Dthermo} F. Mancini, D. Villani and H. Matsumoto,
{\it cond-mat}/9709189.

\bibitem{incom} F. Mancini {\it et al.}, {\it Phys. Rev.
B}, {\bf 57} (1998) 6145; A. Avella {\it et al.}, {\it Phys. Lett. A},
{\bf 240} (1998) 235.

\bibitem{rowe} D. J. Rowe, {\it Rev. Mod. Phys.}, {\bf 40} (1968) 153.

\bibitem{2pole} A. Avella {\it et al.}, {\it Int. J. Mod. Phys. B}, {\bf
12} (1998) 81.

\bibitem{Npole} F. Mancini, {\it cond-mat}/9803276.

\bibitem{bonner} J. Bonner and M. Fisher, {\it Phys. Rev.}, {\bf 135}
(1964) A640.

\bibitem{shibapincus} H. Shiba and P. A. Pincus, {\it Phys. Rev. B},
{\bf 5} (1972) 1966; J. Schulte and M. B\"{o}hm, {\it Phys. Rev. B},
{\bf 53} (1996) 15385.

\end{thebibliography}
\end{document}